\documentclass[%
 reprint,
 amsmath,amssymb,
 aps,
prb,
floatfix,
]{revtex4-2}

\usepackage{hyperref}
\usepackage{chemformula} % Formula subscripts using \ch{}
\usepackage[T1]{fontenc} % Use modern font encodings
\usepackage{multirow}
\usepackage[english]{babel}

\usepackage{xcolor}

\renewcommand{\selectlanguage}[1]{}

\begin{document}

\title[]
  {Controlling Excitons in Quasi-1D Perovskites by Dielectric Screening and Connectivity
}

\author{Kostas Fykouras}
\author{Linn Leppert}
\email{l.leppert@utwente.nl}

\affiliation{MESA+ Institute for Nanotechnology,
University of Twente, Enschede 7500 AE, The Netherlands}

\begin{abstract}
Reducing the dimensionality of metal-halide perovskites enhances quantum and dielectric confinement, enabling tunable excitonic properties. In one dimension, the arrangement of metal-halide octahedra in chains with corner-, edge-, or face-sharing connectivity allows for additional structural flexibility. This not only expands material design possibilities but also reflects quasi-one-dimensional motifs that arise during perovskite formation but are poorly understood. Using first-principles many-body perturbation theory within the $GW$ and Bethe-Salpeter Equation framework, we provide a comprehensive picture of how one-dimensional confinement, octahedral connectivity and dielectric screening affect optical absorption and exciton photophysics in these materials. Our calculations reveal that increasing octahedral connectivity leads to increased exciton binding and complex, anisotropic optical signatures. However, in experimentally synthesized organic-inorganic systems, pronounced dielectric screening effects can reduce exciton binding energies by several hundred meV, altering these trends. These findings offer insights and design principles for excitonic properties, and aid the interpretation of optical experiments on one-dimensional perovskites.
\end{abstract}

\maketitle
\section{Introduction}
Metal-halide perovskites are a class of materials with record-breaking efficiencies in solar cell applications \cite{green_emergence_2014,jung_perovskite_2015}. Among the most attractive properties of these materials are their facile fabrication and immense compositional tunability. Their chemical formula, ABX$_3$, can accommodate a monovalent cation at the A site (e.g., Cs$^{+}$, CH$_3$NH$_3$$^+$, CH(NH$_2$)$^{2+}$), a divalent metal at the B site (e.g., Pb$^{2+}$, Sn$^{2+}$) and a halide at the X site (e.g., I$^-$, Br$^-$, Cl$^-$) \cite{alsalloum_low-temperature_2020}. Dimensional reduction can be achieved by incorporating a larger A-site molecule, yielding quasi-two dimensional (2D) structures in which large aromatic or aliphatic organic spacer cations separate sheets of corner-sharing metal-halide octahedra \cite{smith_between_2017}. These quasi-2D materials exhibit tunable quantum and dielectric confinement effects because their layered sublattice structure resembles a quantum-well heterostructure, leading to excitonic features that persist at room temperature \cite{blancon_semiconductor_2020}. By changing the organic spacer \cite{cao_2d_2015, cheng_extremely_2018, marongiu_role_2019, Filip2022, chakraborty_quantum_2022, simbula_exciton_2023} and the number of inorganic layers $n$ \cite{Blancon2018, Cho2019}, band gaps, exciton binding energies, and other optoelectronic properties of these materials can be tailored for a wide range of applications, from photovoltaics \cite{Tsai2016} to spin-optoelectronics \cite{li_large_2024, chakraborty_design_2024}.

Further enhancement of confinement affects and structural tunability can be achieved in quasi-1D organic-inorganic perovskites. In these materials metal-halide octahedra can arrange with corner-, edge- and face-sharing connectivity \cite{zhang_one-dimensional_2022,yuan_one-dimensional_2017,wong_classification_2021}, and in various geometries, for example, in linear or zigzag chains, depending on octahedral connectivity and the type of organic cation \cite{rahaman_one-dimensional_2021,duan_recent_2023}. Since the first reported quasi-1D perovskite by Mitzi \textit{et al.}~\cite{mitzi_conducting_1995}, these materials have attracted increasing attention. Given their tunable structural, chemical, and photophysical properties, quasi-1D perovskites show potential for a wide range of applications \cite{duan_recent_2023}. For example, in 2018, Mao \textit{et al.} fabricated (2,6-dmpz)$_3$Pb$_2$Br$_{10}$ which contained a network of edge- and corner-sharing octahedra and had a photoluminescence quantum yield of 12\% \cite{mao_structural_2018}. Yuan \textit{et al.} reported C$_4$N$_2$H$_{14}$PbBr$_4$ with an edge-sharing octahedral network which showed strong quantum confinement and the formation of self-trapped excitons \cite{yuan_one-dimensional_2017}. Furthermore, Yu \textit{et al.} observed a high photoluminescence quantum efficiency of 60\% in (C$_4$N$_2$H$_{12}$)$_3$(PbBr$_5$)$_2$·4DMSO that exhibited a quasi-1D  structure with corner-sharing octahedral connectivity \cite{yu_one-dimensional_2021}. Despite their significant potential, experimental and computational research on quasi-1D perovskites remains limited compared to their quasi-2D and 3D counterparts. This limited attention has partly stemmed from the absence of systematic methods for synthesizing well-defined low-dimensional structures. However, recent advances in fabrication and characterization \cite{lin_low-dimensional_2018,chen_critical_2023,Wan2025-sk,Chandra_Patra2025-uh,Quarti2025-ax} have enabled the targeted exploration of the optoelectronic functionalities of these materials \cite{wang_applications_2022,zhang_synthesis_2016}.

Additionally, recent studies have shown that the formation of low-dimensional iodoplumbate intermediates with varying octahedral connectivity during perovskite crystallization can strongly influence film morphology and device efficiency \cite{guo_identification_2016, petrov_crystal_2017, babbeOpticalAbsorptionBasedSitu2020, songRevealingDynamicsHybrid2020, hwangChallengesControllingCrystallization2023, wangControllableIodoplumbateCoordinationHybrid2024}. In-situ optical probes such as UV–vis absorption and photoluminescence (PL) have been proposed to track these intermediates \cite{babbeOpticalAbsorptionBasedSitu2020, zhuIodoplumbateComplexTransformation2025, Spies2025}, although their unambiguous identification remains a major challenge.

Computational studies of quasi-1D halide perovskites have primarily focused on exploring the relationship between their low-dimensional structures, octahedral connectivity, and optoelectronic properties \cite{qian_theoretical_2017, Jiang2022-xq, Xue2022-ye}. In particular, Kamminga \textit{et al.} and Deng \textit{et al.} investigated the relationship between dimensionality, octahedral connectivity and electronic structure using first-principles Density Functional Theory (DFT). These studies showed that increasing connectivity - from corner-sharing (one shared iodine) to edge-sharing (two shared iodines) to face-sharing (three shared iodines) - leads to an increase in the band gap, indicating considerable tunability of band gaps through chemical substitution and confinement effects \cite{Kamminga2016, kamminga_role_2017,Deng2018-hg}.

Despite the insights provided by these computational studies, significant gaps remain in understanding the photophysical properties of quasi-1D perovskites. Such understanding is not only of fundamental interest, but also essential for guiding the design of low-dimensional perovskites for optoelectronic and energy applications. In particular, the interplay of quantum confinement, dielectric environment, and exciton photophysics plays a central role in determining light absorption and charge carrier dynamics in devices such as solar cells and photodetectors. To the best of our knowledge, most recent studies on quasi-1D perovskites rely on DFT, which - as a ground-state theory - neglects the pronounced excitonic effects that dominate the optical response in low-dimensional materials. Excited-state methods, such as Green's function-based many-body perturbation theory in the $GW$ and Bethe-Salpeter Equation (BSE) framework, are essential to accurately capture these excitonic features \cite{rohlfing_electron-hole_2000, Puschnig2002, wirtz_excitons_2006, Qiu2013, Reyes-Lillo2016, giorgi_nature_2018, qiu_signatures_2021, leppert_excitons_2024}. Moreover, while corner-sharing octahedral structures have been associated with smaller band gaps than their edge- and face-sharing counterparts \cite{Gilley2025-di}, recent work has shown that quantum confinement effects \cite{chen_tunable_2023}, organic cation screening \cite{Filip2022}, and orbital interactions between organic and inorganic sublattices \cite{Umeyama2020a, Matheu2022, forde_dielectric_2023, ni_enhanced_2024, boeijeMolecularEngineeringInterlayer2025} significantly influence band gaps and exciton binding energies. These findings highlight the need for systematic investigations that fully account for dielectric screening, structural connectivity, and excitonic effects. 

Here we use the $GW$ and BSE approaches combined with group theory and electrostatic modeling to investigate the electronic structure, optical absorption, and excitonic properties in quasi-1D halide perovskites with corner-, edge-, and face-sharing connectivity. To isolate the effect of octahedral connectivity from other structural characteristics, we created minimal model structures inspired by known quasi-1D perovskite structures. Our calculations reveal that corner-sharing structures exhibit excitonic finestructures and optical absorption onsets qualitatively similar to their 2D and 3D analogues, albeit with significantly enhanced exciton binding energies and symmetry-induced finestructure splitting between excitonic states polarized along and perpendicular to the 1D perovskite chain directions. As connectivity increases from corner- to edge- and face-sharing, group-theoretical analysis indicates substantial changes in the character of low-energy excitations, leading to distinct optical signatures at the absorption onset, accompanied by increases in both band gaps and exciton binding energies. Importantly, however, we observe entirely reversed trends in the exciton binding energies of experimentally synthesized structures, which we find to be rooted in pronounced confinement and dielectric screening effects. Based on these results, we propose material design strategies for systematically tuning the photophysical properties of quasi-1D halide perovskites by controlling octahedral connectivity, organic spacer dimensions, and the dielectric environment provided by the organic moieties.

\section{Methods}
All first-principles DFT calculations were performed using the PBE exchange-correlation functional \cite{Perdew1996-vh} as implemented in the \textsc{quantum espresso} software package \cite{Giannozzi2009-vq,Giannozzi2017-xc}. Norm-conserving fully relativistic pseudopotentials were used from PseudoDojo \cite{Van_Setten2018-mo} with the following electronic configurations: Pb $\mathrm{5d^{10}6s^26p^2}$, I $\mathrm{5s^25p^5}$ and Cs $\mathrm{5s^25p^66s^1}$. Spin-orbit coupling (SOC) was included self-consistently in all calculations except for structural relaxations. For the construction of the zeroth-order single-particle Green's function $G_0$ and screened Coulomb interaction $W_0$ from Kohn-Sham eigenvalues and eigenfunctions, and the solution of the BSE we used the \textsc{berkeleygw} software package, version 4.0 \cite{Deslippe2012-lw, barker_spinor_2022}. A detailed list of computational parameters and convergence tests can be found in the Supplemental Material. Effective masses were obtained by calculating and diagonalizing the effective mass tensor using the Effective Mass Calculator (EMC) program  \cite{githubAfonariOverview}.
\begin{figure}[htb]
  \includegraphics[width=\columnwidth]{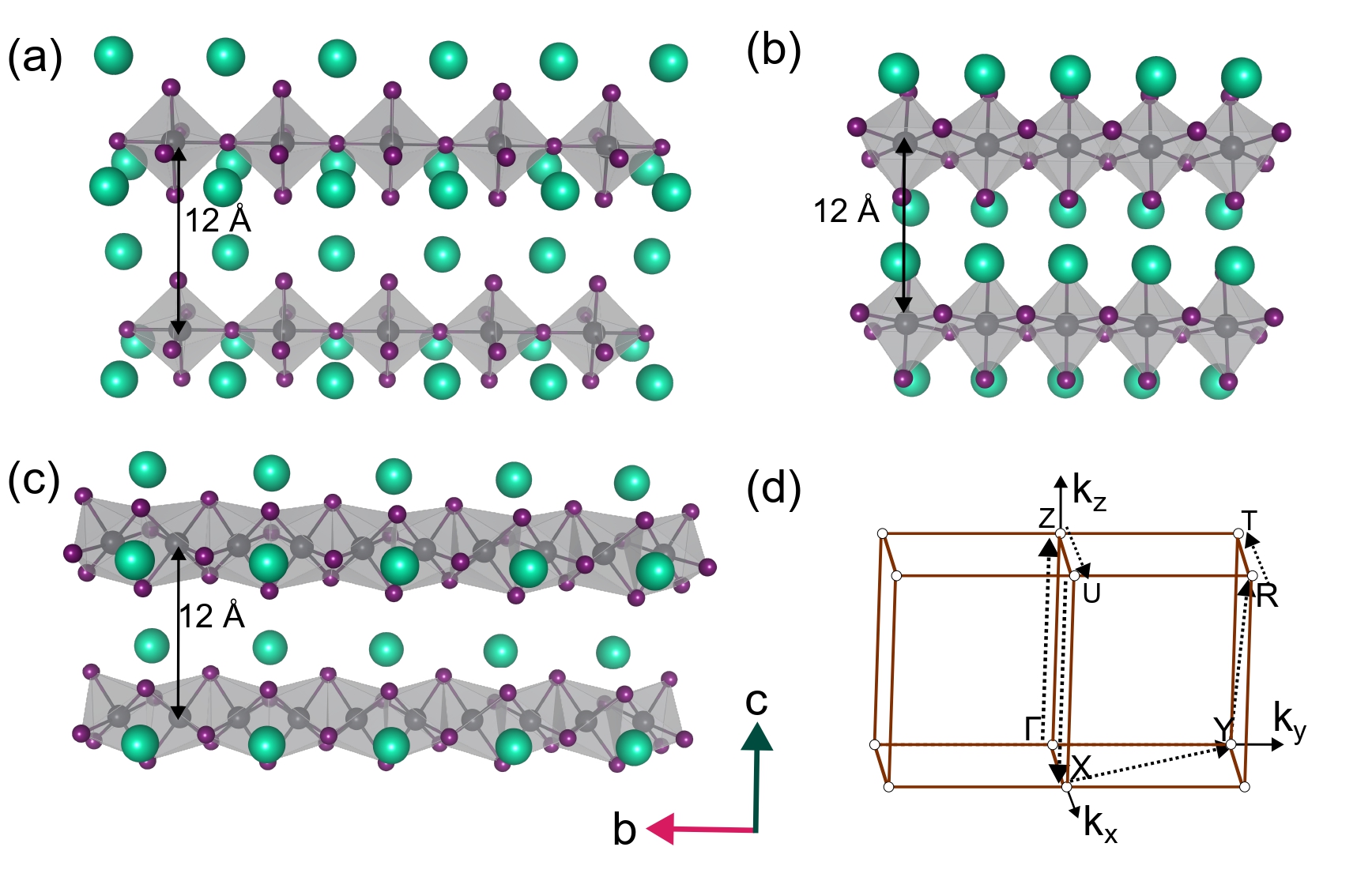}
  \caption{Geometries of model structures with (a) corner-, (b) edge- (c) face-sharing connectivity and (d) the Brillouin zone of the model structures with the path chosen for bandstructure calculations shown as black dotted arrows. The chain-chain distance for these geometries is indicated by the black arrow and is 12\,{\AA} for all model structures. Purple spheres represent I, gray spheres Pb and green spheres Cs atoms.}
  \label{Figure1}
\end{figure}

We used a set of model and experimental structures to disentangle the effects of dimensionality, dielectric screening, octahedral connectivity and specific structural distortions. We started by constructing three sets of model structures with corner-, edge-, and face-sharing octahedral connectivity, shown in Fig.~\ref{Figure1} (a), (b) and (c). The space groups of the model structures are P$_{mm2}$, P$_{2/m}$ and P$_{2_1}$  for the corner-, edge- and face-sharing structure, respectively. We included Cs cations in these structures for charge compensation to avoid the appearance of spurious states below the conduction band minimum (CBM) as reported in Ref.~\citenum{kamminga_role_2017}. The resulting chemical formulas for the corner-, edge- and face-sharing geometry are Cs$_3$PbI$_5$, Cs$_2$PbI$_4$ and Cs$_2$Pb$_2$I$_6$, respectively. The chain direction is [010], i.e., along the b-lattice vector. The corresponding Brillouin zone of the corner-sharing structure is shown in Fig.~\ref{Figure1} (d), where the direction of the 1D chain corresponds to $\Gamma$ to Y in reciprocal space. The direction $\Gamma$ to Y also represents the in-chain direction for the edge- and face-sharing structures. Although the different model structures belong to different space groups and thus have slightly different Brillouin zones, for ease of comparison, we will use the high-symmetry point labels of the P$_{mm2}$  Brillouin zone for all systems.
Details on the construction and relaxation of these model structures and the experimental structures can be found in the Supplementary Material.

\section{Results and Discussion}
\begin{figure}[htb]
  \includegraphics[width=\columnwidth]{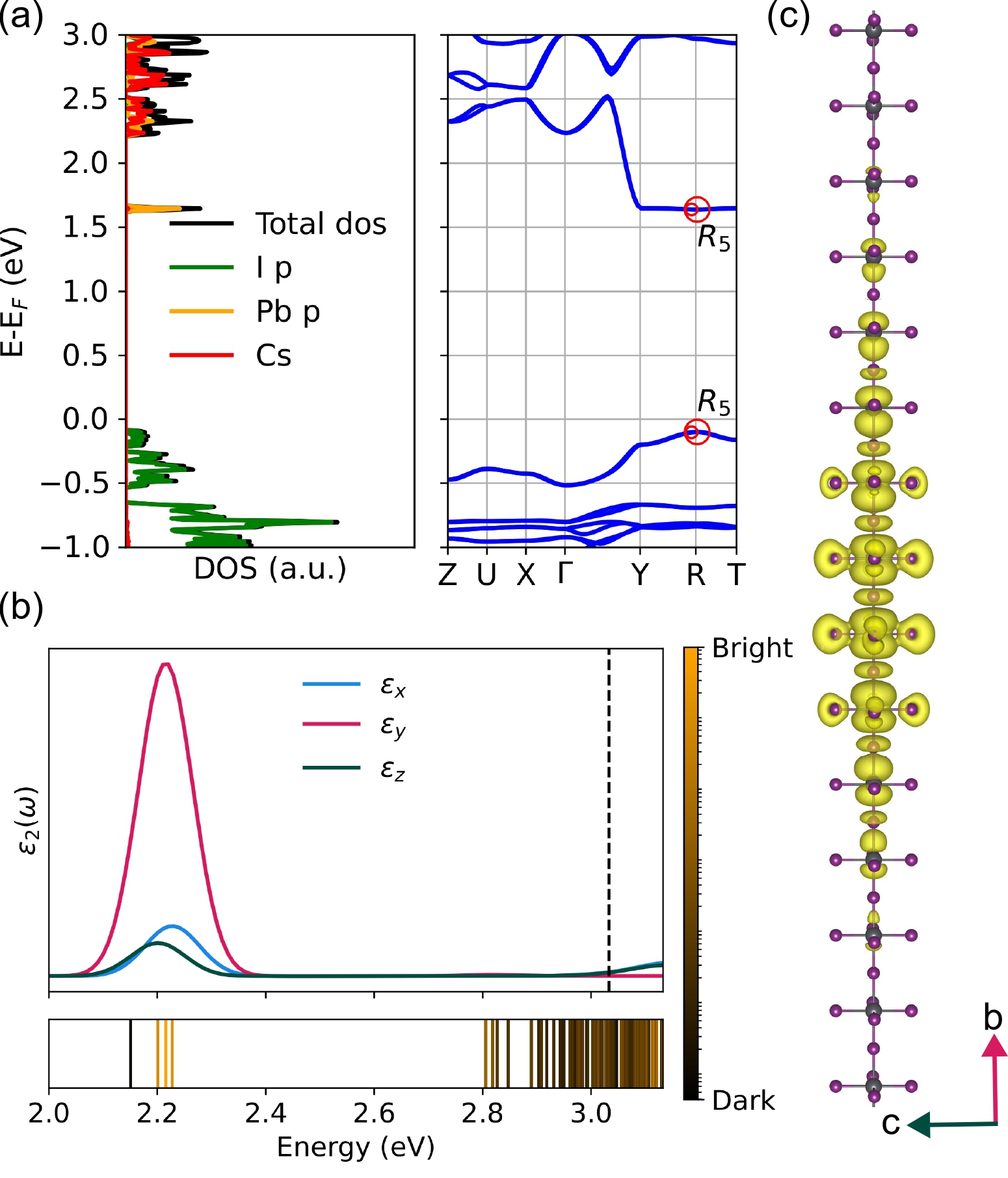}
  \caption{Electronic and optical properties of the corner-sharing structure. (a)  DFT@PBE+SOC DoS (left) and bandstructure (right). Black, green, orange and red lines indicate total, I p, Pb p and Cs contributions to the DoS. Pb s contributions to the VBM are not shown. The bandstructure highlights the irreducible representation (IR) of the orbitals at the high symmetry point R. Open red circles indicate $\mathbf k$-points that contribute to the 1s exciton state (b) Imaginary part of the dielectric function and oscillator strengths of excitonic transitions calculated using G$_0$W$_0$+BSE. The labels $\varepsilon_x$, $\varepsilon_y$ and $\varepsilon_z$ correspond to the dependence of the imaginary part of the dielectric function on the direction of light polarization. The dotted black line indicates the G$_0$W$_0$@PBE+SOC band gap. (c) Exciton wavefunction of the first bright exciton, calculated by placing the hole on one of the I atoms. Cs atoms are omitted for clarity.}
  \label{Figure2}
\end{figure}

The electronic and excitonic properties of the corner-sharing model structure are shown in Fig.~\ref{Figure2}. The projected density of states (DoS), calculated using DFT with the PBE functional and including spin-orbit coupling (SOC) self-consistently (Fig.~\ref{Figure2} (a)) is reminiscent of that of 3D \ch{APbX3} halide perovskites,   with the valence band maximum (VBM) derived from Pb s states and X p states and the conduction band minimum (CBM) comprised of Pb p states. The X p derived states in the VBM arise from the in-chain and out-of-chain (dangling) I(5p) orbitals as seen in other low-dimensional corner-sharing systems \cite{Gao2022-ri}. Contrary to 3D \ch{APbX3} and most known quasi-2D halide perovskites, we observe that electronic states derived from the A-site cation hybridize with the Pb p bands at energies slightly above the CBM, because of Cs-Cs interactions between neighboring periodic cells that are absent in quasi-2D or 3D structures. However, as the chain-chain distance is increased, these Cs-derived states are found at higher energies (Fig.~S1). 

The low-dimensional character of the structure is apparent from the bandstructure where dispersion is observed along the direction of the 1D chain of corner-sharing octahedra and little electronic interaction is present along the directions perpendicular to the chain. We quantify the dispersion of the bands in the VBM and CBM by calculating the effective masses in the direction parallel and perpendicular to the chain. For the corner-sharing structure the effective masses of electrons and holes in the in-chain direction are similar to the ones reported for 3D CsPbI3 and quasi-2D metal-halide perovskites \cite{Shi2023-iz,Fraccarollo2020-ep} with values of m$^*_{h_{\parallel}}$=-0.42m$_e$ for the holes and m$^*_{e_{\parallel}}$=0.14m$_e$ for the electrons, where m$_e$ is the free electron mass (Table~\ref{Tab1}). As expected in the almost dispersionless out-of-chain the effective masses of of both electrons and holes are significantly larger (m$^*_{h_{\perp}}$ $\ll$ -m$_e$ and m$^*_{e_{\perp}}$ $\gg$ m$_e$). The band gap is direct at the high-symmetry point R. As shown in Table~\ref{Tab1}, using a “one-shot” G$_0$W$_0$ approach, in which DFT-PBE+SOC eigenvalues are corrected perturbatively by constructing the zeroth-order self-energy from PBE eigenvalues and eigenfunctions, leads to an opening of the band gap by 1.29\,eV, similar to reports for 3D and quasi-2D perovskites \cite{PhysRevB.90.245145,PhysRevMaterials.3.103803,Filip_2024,Ahmed_2014,doi:10.1021/acs.jpclett.9b02491}. 
\begin{table}[htb]
  \caption{Lowest direct band gaps E$_{DFT}$ and E$_{GW}$ (in eV) calculated with DFT@PBE+SOC and G$_0$W$_0$@PBE+SOC, respectively, exciton binding energy E$_x$ (in meV), static dielectric constant $\epsilon_{\infty}$, exciton extent d$_{in-chain}$ (in {\AA}) in the in-chain direction of the lowest-energy bright state and ratio of oscillator strength of lowest-energy bright (OS$_b$) and dark state (OS$_d$), and calculated effective masses of the in-chain ($m^*_{\parallel}$) direction, in units of free electron mass $m_e$ for the $\mathbf{k}$-points where the VBM and CBM is located for corner-, edge-, and face-sharing model structures and experimental structures.
}
  \label{Tab1}
  \begin{tabular}{lc|c|c|c|c|c}
  & \multicolumn{3}{c|}{Model} & \multicolumn{3}{|c}{Experimental} \\
    \hline
      & Corner & Edge & Face & Corner & Edge & Face  \\
    \hline
    E$_{DFT}$               & 1.73&   1.90 &     2.54 & 1.92 & 2.19 & 1.53     \\
    E$_{GW}$                &3.03 &   3.57 &     4.03  & 2.98 &  3.22 & 3.16   \\
    E$_x$                   & 881 &   1082  &    1276 & 541 & 476 & 416         \\
    $\epsilon_{\infty}$     & 2.57&   2.44   &   2.35  & 3.68 & 3.86 & 4.06     \\
    d$_{in-chain}$          & 32.24 &  20.79  &   20.95 &-- &-- &--       \\
    OS$_b$/OS$_d$           & 1.1x10$^6$ &     7.5x10$^2$ & 2.4x10$^5$ &-- &-- &--     \\
    m$^*_{h_{\parallel}}$     & -0.42 &   -0.95 & -1.01 &-- &-- &--      \\
    m$^*_{e_{\parallel}}$     & 0.14 &    0.40  & 1.20  &-- &-- &--     \\

    \hline
  \end{tabular}
\end{table}

We then use the BSE approach to calculate the imaginary part of the dielectric function, as shown in Fig.~\ref{Figure2} (b), including electron-hole interaction effects to extract exciton binding energies and finestructures (see Supplemental Material for computational details and convergence tests). In line with its 1D nature, the corner-sharing structure has a large exciton binding energy of 881\,meV (Table~\ref{Tab1}) double the value that was reported for the equivalent quasi-2D model structure \cite{https://doi.org/10.1002/adom.202202801}. The imaginary part of the dielectric function has its largest intensity along the direction of the chain. Similar to the procedure described in Ref.~\citenum{https://doi.org/10.1002/adom.202202801,doi:10.1021/acs.jpclett.5b00905,Quarti2020-gd}, we apply group theory to analyze the excitonic finestructure. The high-symmetry $\mathbf k$-point R of the corner sharing geometry belongs to the C$_{2v}$ double point group (Table~S2), in which the VBM and CBM have the irreducible representation (IR) R$_5$. The product of those two IRs with an S-like envelope to represent the first excited state in the Wannier picture, produces four states with IR A$_1$+A$_2$+B$_1$+B$_2$. Given that the dipole operator transforms as A$_1$+B$_1$+B$_2$ in the C$_{2v}$ point group, we ultimately obtain one dark and three bright states. This result is reminiscent of the excitonic fine structure of 3D and 2D halide perovskites \cite{Quarti2020-gd,doi:10.1021/acs.jpclett.5b00905}. However, in our quasi-1D chains, the three bright states, which correspond to excitations with perpendicular transition dipole moments, are non-degenerate due to the 1D crystal field of the structure. This is in line with our BSE results in Fig.~\ref{Figure2} (b), which show that the onset of the imaginary part of the dielectric function is composed of four states, one dark, and three non-degenerate bright corresponding to one in-chain and two out-of-chain directions. 

We follow the procedure described in Ref. \citenum{Sharifzadeh2013-ze} and \citenum{Biega2021-xf} to quantify the spatial extent of the exciton wavefunction $\Psi(\mathbf r_r, \mathbf r_h)$. For that we calculate the average electron-hole separation $\sigma=\sqrt{\langle\textbf{r}^2\rangle-\langle\textbf{r}\rangle^2}$ where $\langle\textbf{r}^n\rangle=\int_{\Omega}\textbf{r}^n F_{s}(\textbf{r})d^3\textbf{r}$ is the $n$’th moment of the electron-hole correlation function $F_{s}(\textbf{r})$ defined as $F_{s}(\textbf{r})=\int_{\Omega}d^3\textbf{r}_h|\Psi(\textbf{r}_e=\textbf{r}_h+\textbf{r},\textbf{r}_h)|^2$. $F_{s}(\textbf{r})$ corresponds to the probability of finding an electron-hole pair separated by a vector $\textbf{r}=\textbf{r}_e+\textbf{r}_h$ in a volume $\Omega$. The corresponding exciton wavefunction, shown in Fig.~\ref{Figure2}(c), has an in-chain spatial extent of 32.24\,{\AA} (Table~\ref{Tab1}) and minimal extension in the out-of-chain direction (Fig.~S2).

\begin{figure*}[htb]
  \includegraphics[width=\textwidth]{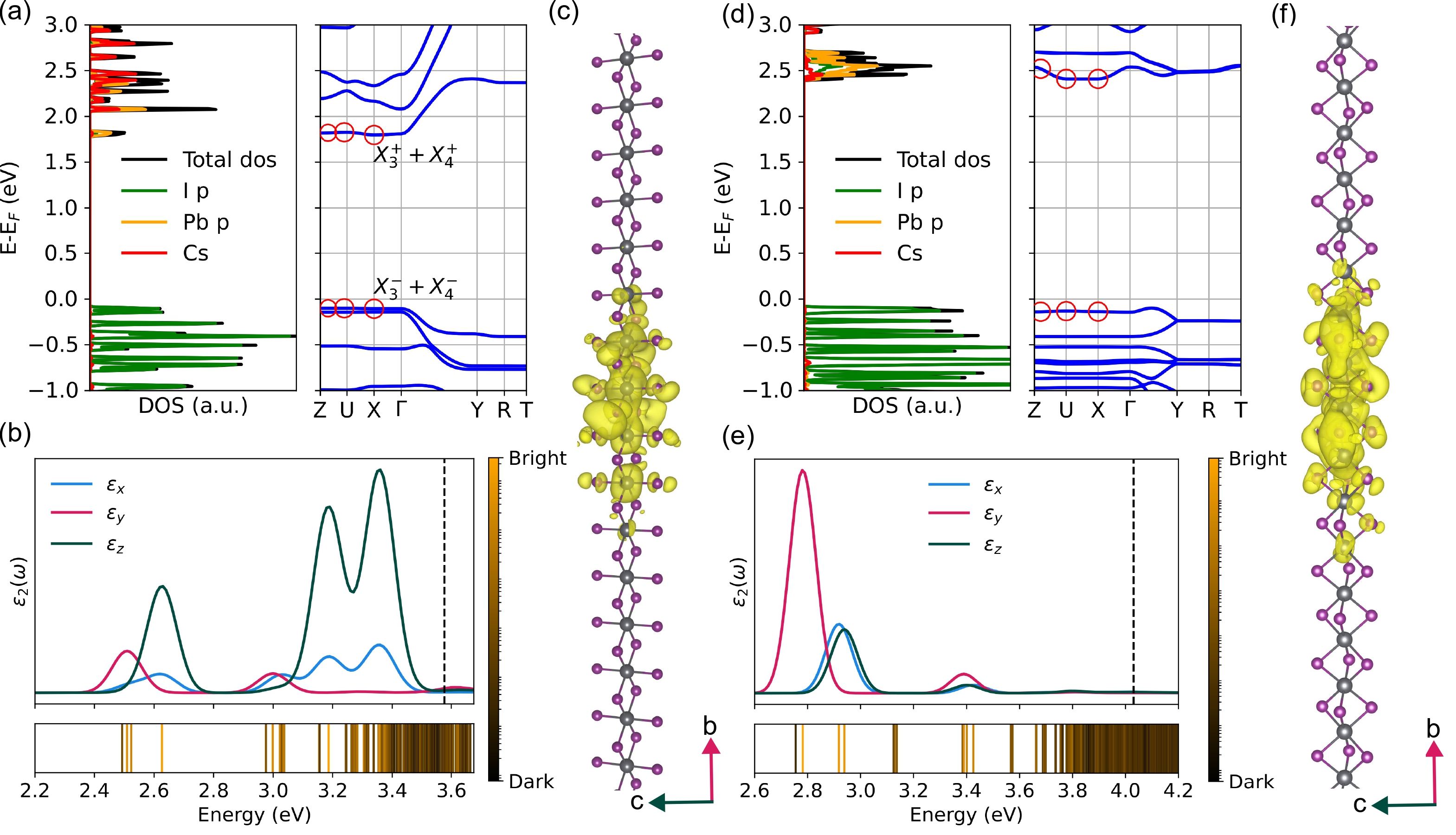}
  \caption{Electronic and optical properties of edge- and face-sharing structures. All labels and colors defined as above. DFT@PBE+SOC projected DoS (left) and bandstructure (right) of the (a) edge-sharing and (d) face-sharing structures. The irreducible representation (IR) of the orbitals at the high-symmetry point X are mentioned where applicable. Open red circles indicate $\mathbf k$-points that contribute to the 1s exciton state. Imaginary part of the dielectric function and oscillator strengths of excitonic transitions of the (b) edge-sharing and (e) face-sharing structures. The dotted black lines indicate the $G_0$W$_0$@PBE+SOC band gap. Exciton wavefunction of the first bright correlated exciton, calculated by placing the hole on one of the I atoms for the (c) edge-sharing and (f) face-sharing structures.}
  \label{Figure3}
\end{figure*}
We now turn to the edge- and face-sharing geometries, shown in Fig.~\ref{Figure3}. The bandstructure of the edge-sharing structure, Fig.~\ref{Figure3}(a), has flat bands in the out-of-chain directions U to X to $\Gamma$, and disperse bands in the in chain $\Gamma$ to Y direction with electron and hole effective masses of m$^*_{e_{\parallel}}$=0.40m$_e$ and m$^*_{h_{\parallel}}$=-0.95m$_e$ respectively (Table~\ref{Tab1}). The lowest direct electronic transition from VBM to CBM occurs at the high-symmetry point X. The position of the band gap is at a different $\mathbf k$-point than in the corner-sharing structure due to stronger orbital hybridization between the I and Pb orbitals at the point R (Fig.~S3) lowering (increasing) the energy of the valence (conduction) band.
The DFT-PBE+SOC (Table \ref{Tab1}) band gap of the edge-sharing model structure is 1.90\,eV, with a G$_0$W$_0$ correction of 1.67\,eV. The exciton binding energy of 1082\,meV is significantly higher than that of the corner-sharing structure.
For the edge sharing geometry the high-symmetry $\mathbf k$-point X belongs to the C$_{2h}$ point group (Table~S2) with IR X$_{3}^{-}$+X$_4^-$ for the VBM and X$_{3}^{+}$+X$_4^+$ for the CBM. Our $GW$+BSE calculations show that the onset of the dielectric function of the edge-sharing structure consists of four bright states; one of which has a two orders of magnitude lower oscillator strength than the other three but is significantly brighter than the lowest-energy dark state in the corner-sharing structure (Table~\ref{Tab1}). 
This result can be explained by examining the product of the IR of the VBM and CBM which produces four states with IR 2A$_u$+2B$_u$ that have odd symmetry. Since the dipole operator transforms as A$_u$+2B$_u$ in the C$_{2h}$ point group, this results in four symmetry-allowed states that make up the onset of the dielectric function (Fig.~\ref{Figure3} (b)).

Fig.~\ref{Figure3} (b) shows that the onset of the imaginary part of the dielectric function originates from the in-chain direction, while, contrary to the corner-sharing geometry, the direction of the highest intensity is the out-of-chain c-direction. This is because in the edge-sharing structure, out-of-chain (axial) I p orbitals contribute significantly more to the VBM than edge-connecting (equatorial) I p orbitals (Fig.~S3). This out-of-chain intensity is also present when the distance between the chains is increased (Fig.~S4). Additionally, we note that excitonic absorption in the edge-sharing geometry has a more complex signature compared to the corner-sharing structure, arising from transitions between the relatively flat bands that comprise the band edges. The exciton wavefunction, shown in Fig.~\ref{Figure3}(c), is much more localized in the in-chain direction as compared to the corner-sharing structure, with an in-chain spatial extent of 20.79 {\AA} (Table~\ref{Tab1}).

In the face-sharing structure, the lowest-energy VBM to CBM transition is located at the high-symmetry point U (Fig.~\ref{Figure3}(d)). Analogous to the two other connectivity motifs, the reciprocal-space direction corresponding to the in-chain direction exhibits the largest electronic dispersion, albeit significantly smaller than that of the corner- and edge-sharing structures with electron and hole effective massses of m$^*_{e_{\parallel}}$=1.20m$_e$ and m$^*_{h_{\parallel}}$=-1.01m$_e$ respectively  (Table~\ref{Tab1}). The band gap, calculated to be 2.54\,eV with DFT-PBE+SOC and increased by 1.48\,eV with G$_0$W$_0$ (Table~\ref{Tab1}), is the largest of the three model structures, in line with earlier reports \cite{kamminga_role_2017}. Consequently, the exciton binding energy of 1276\,meV is larger than that of the corner- and the face-sharing structure too. The imaginary part of the dielectric function (Fig.~\ref{Figure3} (e)) has the highest intensity in the in-chain direction, with a complex excitonic signature arising from the flat bands that comprise the VBM and CBM. Given the low-symmetry space group of the face-sharing structure, the point group of the high-symmetry point U was not identified. Regardless of that, we observe that the onset of the dielectric function is composed of four states, one dark and three bright with relative oscillator strength similar to the corner-sharing structure (Table~\ref{Tab1}). We note that while the exciton binding energy of the first bright state is larger than that of the edge-sharing structure, the in-chain exciton extent is similar (Table~\ref{Tab1}).

Next, we calculated the optoelectronic properties of experimentally synthesized quasi-1D structures with corner-, edge-, and face-sharing octahedral connectivity, comparing them to our previously discussed model systems. For this comparison, we select three experimentally realized materials: corner-sharing (\ch{(HSC(NH2)2)3PbI5}) \cite{Daub2021-uu}, edge-sharing  (\ch{(C6H10N2)(PbI4)2H2O}) \cite{lemmerer_two_2006}, and face-sharing (\ch{(C2H7N2)6PbI3}) \cite{https://doi.org/10.1002/smll.202403685} shown in Fig.~S5. The experimental structures are optimized with DFT-PBE keeping the inorganic Pb-I network and the lattice vectors fixed, allowing only the molecular A site to relax. Then we calculate the DFT@PBE+SOC and $G_0W_0$@PBE+SOC band gaps and the $GW$+BSE exciton binding energies as before (see Table~\ref{Tab1}). We find that while the corner-sharing geometry has the smallest $GW$ band gap, similar to our calculations for the model structures, the energetic ordering of the band gaps of the edge- and face-sharing structures is reversed: the face-sharing structure's band gap is 80\,meV smaller than that of the edge-sharing structure (Table~\ref{Tab1}). Additionally, we find that the exciton binding energies unexpectedly show an inverted trend as compared to the model structures. Specifically, the face-sharing structure exhibits the smallest exciton binding energy, followed by the edge- and the corner-sharing structure. Moreover, the exciton binding energies of the edge- and face-sharing experimental structures are more than a factor of two smaller than those predicted for the corresponding model systems.
\begin{figure*}[htb]
  \includegraphics[width=\textwidth]{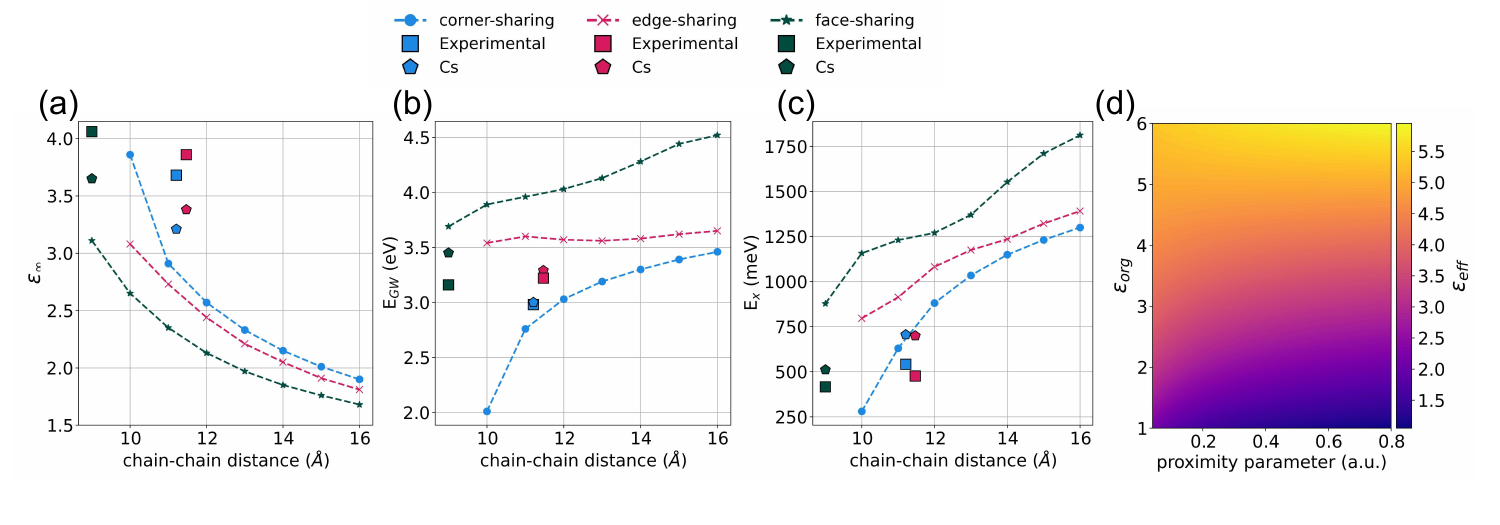}
  \caption{
  Relation between the chain-chain distance in {\AA} and the (a) the static dielectric constant, (b) the E$_{GW}$ band gap and (c) the exciton binding energy E$_x$, for the corner-sharing (blue circles), edge-sharing (red crosses), and face-sharing (green stars) model structures. Lines are meant as guides to the eye. Results for the experimental (Cs-based) structures are represented by squares (pentagons). 
  (d) Contour plot showing the dependence of the effective dielectric constant $ \epsilon_{\text{eff}}$ on $\epsilon_{\text{org}}$ (for a fixed $\epsilon_{\text{inorg}}=5$) and the dimensionless proximity parameter (see text).}
  \label{Figure4}
\end{figure*}

To understand the origin of these reversed trends, we systematically studied how the chain-chain distance and dielectric screening by the organic moieties influence the static dielectric constant ($\epsilon_{\infty}$), the $GW$ band gap, and the exciton binding energy ($E_x$), as shown in Figures~\ref{Figure4}(a)-(c). For this, we calculated the dependence of these quantities on chain-chain distance using the previously discussed model structures. Figure~\ref{Figure4}(a) demonstrates that $\epsilon_{\infty}$ decreases as the chain-chain distance increases, resulting in a monotonic increase in exciton binding energies for all three geometries (Figure~\ref{Figure4}(c)). For the corner- and face-sharing geometries, this increase closely follows the expected $1/\epsilon_{\infty}$ and the $1/\epsilon_{\infty}^2$ relationship for the band gap and the exciton binding energy (Fig.~S6 and~S8). Deviations are observed for the edge-sharing geometry which arise from more complex trends in its $GW$ band gaps (Fig.~S7), reflecting competing effects: reduced dielectric screening at larger distances tends to increase the $GW$ band gap, while enhanced orbital hybridization between Pb-derived conduction-band states and Cs-derived states lowers the conduction band minimum (Fig.~S9).

Additionally, Fig.~\ref{Figure4} includes the three experimental structures (denoted "Experimental") and three equivalent structures in which the molecular A site was replaced by Cs (denoted "Cs"), to isolate the effect of organic spacer screening. In the corner-sharing case, the results for the experimental structure are close to the predictions of the model structure at an equivalent chain-chain difference, indicating minimal structural distortion and only slight differences upon substituting the molecular A site with Cs. This similarity arises primarily due to the inherently small dielectric constant contribution of the molecular site in this case. In contrast, the edge-sharing experimental structure substantially deviates from its model counterpart at the same chain-chain distance, exhibiting a significantly smaller band gap (by about 0.3\,eV) and an exciton binding energy reduced by roughly 400\,meV. This discrepancy partly results from enhanced dielectric screening by the molecular A site. Indeed, replacing this molecular site with Cs notably reduces the dielectric constant, increasing both the band gap and exciton binding energy. The face-sharing case is similar to the edge-sharing scenario. Replacing the molecular A site with Cs leads to a significant increase of the band gap and the exciton binding energy, but does not fully explain the deviation from the results for the model structures. Since the Pb-I bond lengths in these two structures are by construction the same as in the corresponding models, the remaining differences can be attributed to differences in symmetry between model and experimental structures (see SI).

Finally, to predict how the dielectric properties of the organic and inorganic sublattices influence the overall dielectric response of these 1D materials, and thus their band gaps and exciton binding energies, we constructed a simple 2D electrostatic model consisting of circular inorganic regions with dielectric constant $\epsilon_{\mathrm{inorg}}$ and radius $r$ and distance $L$, embedded in a medium representing the organic sublattice with dielectric constant $\epsilon_{\text{org}}$. Chain-chain distance is represented by a dimensionless proximity parameter $(L-2r)/L$, since this simple model only reproduces the exponential decay of the effective static dielectric constant but not the microscopic details at the relevant experimental chain-chain distances (see Supplemental Material for details). Figure~\ref{Figure4}(d) shows that increasing $\epsilon_{\text{org}}$ significantly enhances the effective dielectric constant, highlighting that the organic sublattice plays a key role in determining the dielectric screening. In contrast, varying $\epsilon_{\mathrm{inorg}}$ has a significantly smaller effect (Fig.~S10), in line with observations made for 2D metal-halide perovskites \cite{Filip2022}. These findings underline the importance of dielectric screening of excitons by the organic sublattice when evaluating material candidates for energy applications, where control over exciton binding and dissociation is crucial for optimizing device performance.

\section{Conclusion}
In conclusion, we performed a systematic first-principles study exploring how one-dimensional confinement, octahedral connectivity and dielectric screening influence the optoelectronic properties of quasi-1D metal-halide perovskites. Our results demonstrate that increased sharing of halides in the octahedral chains enhances exciton binding energies and introduces anisotropic and complex optical signatures. However, comparison with experimental structures shows significant deviations from predictions based solely on idealized models, highlighting the importance of confinement (chain-chain distance) and dielectric screening effects. In particular, our electrostatic model underscores that the dielectric properties of the organic A-site, or adjacent solvent molecules, play a critical role in determining exciton binding energies and optical absorption. These findings indicate clear pathways for tailoring material properties through targeted molecular and structural design, particularly in the context of optoelectronic and light-harvesting materials for energy technologies. 

Our findings also have implications for the ongoing debate about the role of iodoplumbate complexes and other low-dimensional motifs during perovskite film formation. In-situ optical probes such as UV–vis absorption and PL have been employed to track solvated PbI$_2$ clusters in solution \cite{Spies2025} and the transformation of iodoplumbate species in the solid state \cite{manserEvolutionOrganicInorganic2015, zhuIodoplumbateComplexTransformation2025}, suggesting signatures of intermediate octahedral connectivity. Complementary techniques, including GIWAXS, Raman, and EXAFS, have been used to help resolve structural motifs during growth \cite{pratapOutofequilibriumProcessesCrystallization2021, hwangChallengesControllingCrystallization2023}. More recently, nuclear quadrupolar resonance spectroscopy has been proposed as a powerful probe of local halide environments, with sensitivity to different octahedral connectivities \cite{Quarti2025-ax}. Our calculations demonstrate how excitonic properties vary systematically with connectivity and dielectric screening and thus provide testable predictions for identifying such motifs in real-time studies. Combining polarization-resolved in-situ optical spectroscopy with structural probes may ultimately allow for a more reliable mapping between transient intermediates and their excitonic fingerprints, thereby informing strategies to control crystallization pathways and optimize film quality.

\begin{acknowledgments}
The authors acknowledge funding from the Dutch Research Council (NWO) under grant number OCENW.M20.337 and computational resources provided by the Dutch National Supercomputing Center Snellius supported by the SURF cooperative. We thank M. Kamminga and R. Havenith for providing the model structures used in Ref.~\citenum{kamminga_role_2017} and R. Kerner for insightful discussions and comments on an earlier version of this manuscript.
\end{acknowledgments}

%\bibliography{references}

\end{document}